\documentclass[a4paper,10pt,twoside]{cpc-hepnp}
\usepackage{CJK,upgreek,fancyhdr}
\usepackage{multicol}
\usepackage{graphicx}
\usepackage{amssymb,bm,mathrsfs,bbm,amscd}
\usepackage[tbtags]{amsmath}
\usepackage{lastpage}
\usepackage{longtable}
\usepackage{changes}
\usepackage{etoolbox}
\usepackage{setspace}
\usepackage{tabularx}
\usepackage{lineno,hyperref}
\usepackage{supertabular}
\usepackage[normalem]{ulem}
\usepackage{eqnarray}
\usepackage{booktabs}  
\usepackage{subfigure}
\usepackage{float} 
\begin{document}
\begin{CJK*}{GB}{gbsn}

\fancyhead[c]{\small Chinese Physics C~~~Vol. xx, No. x (202x) xxxxxx}
\fancyfoot[C]{\small 010201-\thepage}

\title{Systematic study of proton radioactivity half-lives based on the relationship between the Skyrme-Hartree-Fock and the macroscopic quantities of nuclear matter
\thanks{This work is supported in part by the National Natural Science Foundation of China (Grant No. 12175100), the construct program of the key discipline in Hunan province, the Research Foundation of Education Bureau of Hunan Province, China (Grant No. 18A237), the Natural Science Foundation of Hunan Province, China (Grant Nos. 2015JJ3103, 2015JJ2123), the Innovation Group of Nuclear and Particle Physics in USC.}}

\author{%
Jun-Hao Cheng (程俊皓)$^{1}$
\quad Zhen Zhang (张振)$^{1}$
\quad Xi-Jun Wu (吴喜军) $^{1}$
\quad Peng-Cheng Chu (初鹏程)$^{1}$
\quad Xiao-Hua Li (李小华)$^{1,3,4;1)}$\email{lixiaohuaphysics@126.com}%
}
\maketitle
\address{%
$^1$ School of Nuclear Science and Technology, University of South China, Hengyang 421001, China\\
$^2$ Sino-French Institute of Nuclear Engineering and Technology, Sun Yat-sen University, Zhuhai 519082, China\\
$^3$ School of Math and Physics, University of South China, Hengyang 421001, China\\
$^4$ School of Science, Qingdao Technological University, Qingdao 266000, China\\
$^5$ Cooperative Innovation Center for Nuclear Fuel Cycle Technology \& Equipment, University of South China, Hengyang 421001, China\\
$^6$ Key Laboratory of Low Dimensional Quantum Structures and Quantum Control, Hunan Normal University, Changsha 410081, China\\
}
\begin{abstract}
In the present work, we systematically study the proton radioactivity half-lives of 33 spherical nuclei based on the relationship between the Skyrme parameters and the macroscopic quantities of nuclear matter. Using the two-potential approach with the spherical Skyrme-Hartree-Fock model, the correlation between proton radioactivity half-life and macroscopic quantities was analyzed. Moreover, we obtained a new Skyrme parameter set by fitting the two most weighted macroscopic quantities. Compared with Skyrme parameters MSL0 and the theoretical model of proton radioactivity UDLP, the theoretical proton radioactivity half-life calculated by the new Skyrme parameter set can better reproduce the experimental data.
\end{abstract}


\begin{pacs}
{23.60.+e,} {21.10.Tg}
\end{pacs}

\footnotetext[0]{\hspace*{-3mm}\raisebox{0.3ex}{$\scriptstyle\copyright$}2020
Chinese Physical Society and the Institute of High Energy Physics
of the Chinese Academy of Sciences and the Institute
of Modern Physics of the Chinese Academy of Sciences and IOP Publishing Ltd}%
\begin{multicols}{2}
\section{Introduction}

Proton radioactivity is one of the typical decay mode for odd-Z emitters beyond the proton drip line and is the only decay mode for some intermediate-mass nuclei \cite{SONZOGNI20021}. Since its first discovery in 1970 \cite{JACKSON1970281}, proton radioactivity has become a powerful tool for studying the nature and structure of proton-rich nuclei, as it can provide information about shell structure, bound and unbound nuclear states, \emph{etc.} \cite{KARNY200852}. In addition, the proton radioactivity energy and orbital angular momentum carried by the emitted proton have a significant effect on the half-life of the proton radioactivity \cite{SONZOGNI2002837}. Therefore, the proton radioactivity studies help to determine the orbital angular momentum carried by the emitted proton and characterize its wave function within the nucleus \cite{PhysRevC.69.054311, FERREIRA2007418, PhysRevC.79.054330,2009-0067, Hong_Fei_2009}.

Up to now, there are many methods or models used to study the proton radioactivity half-life can be classified into two categories \cite{Routray2012}. One involves quantum mechanical tunneling through the nuclear mean-field of single-particle resonances \cite{DELION2006113,PhysRevC.58.3011, FERREIRA2011508, doi:10.1063/1.2827286}. Typical methods or models are the non-relativistic Hartree-Fock (HF) calculation using effective phenomenological interactions and covariant density functional theory (CDFT) in the relativistic mean-field (RMF) form \cite{doi:10.1063/1.2827286, PhysRevC.5.626}. Another is the calculation of the probability of proton penetration into the potential barrier by the Wentzel-Kramers-Brillouin (WKB) approximation \cite{Zdeb2016,Chen_2019}, such as two-potential approach (TPA) \cite{PhysRevLett.59.262}, the single-folding model (SFM) \cite{Qian_2010,PhysRevC.72.051601}, the Coulomb potential and proximity potential model (CPPM) \cite{Deng2019, PhysRevC.96.034619}, and so on. In the study of proton radioactivity, the nuclear potential between emitted proton and daughter nucleus  usually determines the accuracy of the model and calculations. In our previous study of proton radioactivity half-life, we successfully described the emitted proton-daughter nucleus nuclear potential using the Skyrme-Hartree-Fock (SHF) microscopic model \cite{CHENG2020121717, cheng2020systematic}. Furthermore, we compared 115 sets of Skyrme interaction parameters constructed for different purposes and found that the most suitable Skyrme parameter to describe the proton radioactivity half-life is SLY7. The construction of the Skyrme interaction parameters is mainly for reasonably describing the ground state properties of nuclei on the periodic table of elements and the saturation properties of symmetric nuclear matter (SNM) rather than for the description of the proton radioactivity process of the nucleus. Therefore, all work using Skyrme-Hartree-Fock to study proton radioactivity may requires a set of Skyrme parameters specifically for calculating proton radioactivity instead of the existing Skyrme parameters. In this work, based on the relationship between the Skyrme parameter and the amount of macroscopic nuclear matter, we use the two-potential approach with Skyrme-Hartree-Fock (TPA-SHF) to study the sensitivity of proton radioactivity half-life to the amount of macroscopic nuclear matter  \cite{PhysRevC.82.024321}. Moreover, we give a set of Skyrme parameters for the study of proton radioactivity based on the constraints on the macroscopic quantities of nuclear matter.

This article is organized as follows. In Sec. II the theoretical framework for proton radioactivity half-life is described in detail. In Sec. III, the detailed calculations, discussion and predictions are provided. A brief summary is given in Sec. IV.

\section{Theoretical framework}
\subsection{Two-potential approach}

The proton radioactivity half-life $T_\frac{1}{2}$ can be expressed by the decay width $\Gamma$, which can be written as 
\begin{equation}
\label{eq1}
T_\frac{1}{2}=\frac{ln2\hbar}{\Gamma}.
\end{equation}
In the framework of the TPA \cite{PhysRevLett.59.262}, the $\Gamma$ depends on the formation probability of the proton radioactivity $S_p$, the normalized factor \emph{F} and the penetration probability of the emitted proton crossing the barrier $P$. It is given by
\begin{equation}
\label{eq2}
\Gamma=\frac{\hbar^2 S_p F P}{4\mu},
\end{equation}
where $\hbar$ is the reduced Planck constant. And $\mu=\frac{M_d M_p}{(M_d+M_p)}$ is the reduced mass, with $M_p$ and $M_d$ being the masses of the emitted proton and the daughter nuclei.

In the classical WKB approximation, the penetration probability $P$ and the normalized factor $F$ are given by
\begin{equation}
\label{eq3}
P=\exp\! [-2 \int_{r_2}^{r_3} k(r)\, dr],
\end{equation}
\begin{equation}
\label{eq4}
F\! \int_{r_1}^{r_2}\frac{1}{2k(r)}\, dr=1.
\end{equation}
where the $r_1$, $r_2$ and $r_3$ represent the classical turning points, they satisfy the conditions $V(r_1)=V(r_2)=V(r_3)=Q_p$. $k(r)$ is the wave number, which can be written as
\begin{equation}
\label{eq5}
k(r)=\sqrt{\frac{2\mu}{\hbar^2}\left|Q_p-V(r)\right|},
\end{equation}
where $Q_p$ is the proton radioactivity energy. The total potential $V(r)$ is given by
\begin{equation}
\label{eq6}
V(r)=V_N(r)+V_C(r)+V_l(r),
\end{equation}
where $V_N(r)$, $V_C(r)$ and $V_l(r)$ represent the nuclear, Coulomb, and centrifugal potentials, respectively. For centrifugal potential $V_{l}(r)$, $l(l + 1) \rightarrow (l + 1/2)^2$ is an essential correction \cite{doi:10.1063/1.531270}. In this work, the centrifugal potential $V_{l}(r)$ is chosen as the Langer-modified form, which can be written as
\begin{equation}
\label{eq9}
V_{l}(r)=\frac{\hbar^2(l+\frac{1}{2})^2}{2{\mu}r^2}, 
\end{equation}
where $l$ is the orbital angular momentum taken away by the emitted proton \cite{PhysRevC.82.059902}. It can be obtained by the parity and angular momentum conservation laws. The Coulomb potential $V_C(r)$ is taken as the potential of a uniformly charged sphere connected to the sharp radius $R=1.28A^{1/3}-0.76+0.8A^{-1/3}$ with $A$ being the mass numbers of parent nucleus \cite{Royer_2000}. It can be expressed as
\begin{equation}
\label{eq7}
\
V_C(r)=\left\{\begin{array}{llll}

\frac{Z_{p}Z_de^2}{2R}[3-(\frac{r}{R})^2], &r<R, \\

\frac{Z_{p}Z_de^2}{r}, &r>R, 

\end{array}\right.
\end {equation}
where $Z_p$ and $Z_d$ are the proton numbers of emitted proton and daughter nuclei, respectively.

\subsection{Spherical Skyrme-Hartree-Fock}

The emitted proton--daughter nucleus nuclear potential $V_N(r)=U_q(\rho, \rho_p, \textbf{\emph{p}})$ is calculated by SHF. The nuclear effective interactions with zero-range, momentum and density dependent forms is given by \cite{PhysRevC.5.626}
\begin{eqnarray}
\label{eq2}
V^{Skyrme}_{12}(\textbf{\emph{r}}_1, \textbf{\emph{r}}_2)=&t_0(1+x_0P_\sigma)\delta(\textbf{\emph{r}})\nonumber\\
&+ \frac{1}{2}t_1(1+x_1P_\sigma)[\textbf{\emph{P}}'^2\delta(\textbf{\emph{r}})+\delta(\textbf{\emph{r}})\textbf{\emph{P}}^2]
\nonumber\\
&+ t_2(1+x_2P_\sigma)\textbf{\emph{P}}'\cdot\delta(\textbf{\emph{r}})\textbf{\emph{P}}
\nonumber\\
&+ \frac{1}{6}t_3(1+x_3P_\sigma)[\rho(\textbf{\emph{R}})]^{\alpha}\delta(\textbf{\emph{r}})
\nonumber\\
&+ i {W_0} \bm{\sigma}\cdot[\textbf{\emph{P}}'\times\delta(\textbf{\emph{r}})\textbf{\emph{P}}], 
\end{eqnarray}
where $\textbf{\emph{r}}=\textbf{\emph{r}}_1-\textbf{\emph{r}}_2$ and $\textbf{\emph{R}}=(\textbf{\emph{r}}_1+\textbf{\emph{r}}_2)/2$, $\textbf{\emph{r}}_i$ (i=1, 2) is the coordinate vector of the $i$-$th$ nucleon. $P_\sigma$ is the spin exchange operator. $\textbf{\emph{P}}'$ and $\textbf{\emph{P}}$ are the relative momentum operators acting on the left and right, respectively. And $t_0$, $t_1$, $t_2$, $t_3$, $x_0$, $x_1$, $x_2$, $x_3$, $W_0$ and $\alpha$ are the Skyrme parameters. 

In local density approximation, the single-nucleon potential from SHF model can be expressed as \cite{PhysRevC.98.054614}

\begin{equation}
\label{eq3}
U_q(\rho, \rho_q, \textbf{\emph{p}})=a\textbf{\emph{p}}^2+b, 
\end{equation}
where $\textbf{\emph{p}}$ is the momentum of the nucleon. The coefficient $a$ and $b$ are given by

\begin{eqnarray}
\label{eq4}
a=&\frac{1}{8}[t_1(x_1+2)+t_2(x_2+2)]\rho\nonumber\\
&+ \frac{1}{8}[-t_1(2x_1+1)+t_2(2x_2+1)]\rho_q, 
\end{eqnarray}

\begin{eqnarray}
\label{eq5}
b=&\frac{1}{8}[t_1(x_1+2)+t_2(x_2+2)]\frac{k^5_{f, n}+k^5_{f, p}}{5\pi^2}\nonumber\\
&+\frac{1}{8}[t_2(2x_2+1)-t_1(2x_1+1)]\frac{k^5_{f, q}}{5\pi^2}
\nonumber\\
&+ \frac{1}{2}t_0(x_0+2)\rho-\frac{1}{2}t_0(2x_0+1)\rho_q
\nonumber\\
&+ \frac{1}{24}t_3(x_3+2)(\alpha+2)\rho^{(\alpha+1)}
\nonumber\\
&-\frac{1}{24}t_3(2x_3+1)\alpha\rho^{(\alpha-1)}(\rho^2_n+\rho^2_p)
\nonumber\\
&-\frac{1}{12}t_3(2x_3+1)\rho^\alpha\rho_q, 
\end{eqnarray}
where $k_{f, q}=(3\pi\rho_q)^{1/3}$ is the Fermi momentum of the nucleon. $\rho_q$ is the proton (neutron) density with $q=p$ $(n)$, $\rho=\rho_p+\rho_n$ represent the total nucleon density.

The total energy $E$ of a nucleon in a nuclear medium can be written as
\begin{eqnarray}
\label{eq6}
E=&U_q(\rho, \rho_q, \textbf{\emph{p}})+\frac{\textbf{\emph{p}}^2}{2m}\nonumber\\
&=\frac{\textbf{\emph{p}}^2}{2m^{*}}+b, 
\end{eqnarray}
where $m^{*}$ stands for effective mass defined by 
\begin{equation}
\label{eq7}
\frac{1}{2m^{*}}=\frac{1}{2m}+a.
\end{equation}
We assume that the total energy of the emitted protons remains constant during the proton radioactivity. $\left|\textbf{\emph{p}}\right|$ can be derived from the total energy, the nucleon density and isospin asymmetry. It can be given by

\begin{equation}
\left|\textbf{\emph{p}}\right|=\sqrt{2m^{*}(E-b)}, 
\end{equation}
and the potential energy $U_q(\rho, \rho_p, \textbf{\emph{p}})=E-\frac{\textbf{\emph{p}}^2}{2m}$.

By comparing expressions in the SHF with corresponding expressions in modified Skyrme-like (MSL) model \cite{PhysRevC.80.014322}, the nine parameters $\sigma$, $\beta$, $\gamma$, $C$, $D$, $y$, $E^{loc}_{sym}(\rho_0)$, the gradient $G_s$ and the symmetry-gradient coefficients $G_v$ in the MSL model can be related to the nine Skyrme interaction parameters by following analytic relations \cite{PhysRevC.82.024321}:
\begin{equation}
t_0=4 \sigma / 3 \rho_0, 
\end{equation}
\begin{equation}
t_1=20C /(9 \rho_0 (k^0_F)^2) +8 G_s/3, 
\end{equation}
\begin{equation}
t_2=\frac{4 (25 C -18 D)}{9 \rho_0 (k^0_F)^2}-\frac{8 (G_s+2 G_v)}{3},
\end{equation}
\begin{equation}
t_3=16 \beta / (\rho^{\gamma}_0(\gamma+1)),
\end{equation}
\begin{equation}
x_0=3(y-1)E^{loc}_{sym}(\rho_0)/\sigma -1/2,
\end{equation}
\begin{equation}
x_1=\frac{12G_v-4G_s-6D/(\rho_0 (k^0_F)^2)}{3 t_1},
\end{equation}
\begin{equation}
x_2=\frac{20G_v+4G_s-5(16C-18D)/( 3 \rho_0 (k^0_F)^2)}{3 t_2},
\end{equation}
\begin{equation}
x_3=-3y(\gamma+1)E^{loc}_{sym}(\rho_0)/(2\beta)-1/2,
\end{equation}
\begin{equation}
\alpha=\gamma-1,
\end{equation}
with $k^0_F=(1.5\pi^2\rho_0)^{1/3}$. Since the seven parameters $\sigma$, $\beta$, $\gamma$, $C$, $D$, $y$, $E^{loc}_{sym}(\rho_0)$ in the MSL model can be expressed analytically in terms of the seven macroscopic quantities i.e. the normal density $\rho_0$, the equation of state at $\rho_0$ $E_0(\rho_0)$, the incompressibility $K_0$, the isoscalar effective mass $m^*_{s,0}$, the isovector effective mass $m^*_{v,0}$, the nuclear symmetry energy at $\rho_0$ $E_{sym}(\rho_0)$ and the extracted density slope $L$ \cite{PhysRevC.80.014322}. The nine Skyrme interaction parameters $t_0$, $t_1$, $t_2$, $t_3$, $x_0$, $x_1$, $x_2$, $x_3$ and $\alpha$ can also be expressed analytically in terms of the nine macroscopic quantities $\rho_0$, $E_0(\rho_0)$, $K_0$, $m^*_{s,0}$, $m^*_{v,0}$, $E_{sym}(\rho_0)$, $L$, $G_S$ and $G_V$ via the above relations.

\section{Results and discussion}
\label{section 3} 
The aim of this work is to study the proton radioactivity half-life based on the relationship between macroscopic quantities and Skyrme parameters \cite{PhysRevC.82.024321}. For study the sensitivity of proton radioactivity to macroscopic quantities \cite{PhysRevC.82.024321}, we adopt the set of Skyrme parameters MSL0 obtained by empirical values of macroscopic quantities to ensure that the adjustment of macroscopic quantities is not out of line with the actual values. The macroscopic quantities and the corresponding Skyrme parameters are listed in Tab. \ref{table 1}.  $G'_0(\rho_0)$ is the Landau parameter whose value can vary from approximately 0 to 1.6, depending on the method and the model \cite{BORZOV2006645,PhysRevC.72.067303,PhysRevC.72.014310}. In this work, the experimental data of spin, parity, the proton radioactivity energy $Q_p$, and proton radioactivity half-lives are taken from the latest evaluated nuclear properties table NUBASE2020 \cite{NUBASE2020} and the latest evaluated atomic mass table AME2020 \cite{CPC-2021-0034,CPC-2020-0033} except for those of $^{140}$Ho, $^{144}$Tm, $^{151}$Lu, $^{159}$Re, and $^{164}$Ir, which are taken from Ref. \cite{BLANK2008403}. The standard deviation $\Delta$ indicates the divergence between experimental data and the the proton radioactivity half-lives calculated by TPA-SHF, which can be expressed as $\Delta=\sqrt{\sum ({\rm{lg}{T^{\rm{exp}}_{1/2}}(s)}-{\rm{lg}{T^{\rm{cal}}_{1/2}}(s)})^2/n}$.

\begin{table*}[!hbt]
\centering
\caption{Macroscopic quantities and corresponding Skyrme parameters in MSL0.}
\label{table 1}
\renewcommand\arraystretch{1.5}
\setlength{\tabcolsep}{2.5pt}
\setlength\LTleft{-13in}
\setlength\LTright{-13in plus 1 fill}
\begin{tabular}{cccc}
\hline
\multicolumn{1}{c}{Quantity}&MSL0&$\rm{Quantity}$&MSL0\\
\hline
$\rho_0$&0.16 $fm^{-3}$&$t_0$&-2118.06 MeV $fm^{5}$\\
$E_0(\rho_0)$&-16MeV&$t_1$&395.196 MeV $fm^{5}$\\
$K_0$&230MeV&$t_2$&-63.953 MeV $fm^{5}$\\
$m^*_{s,0}/m$&0.8&$t_3$&12857.7 MeV $fm^{3+3\alpha}$\\
$m^*_{v,0}/m$&0.7&$x_0$&-0.0709496\\
$E_{sym}(\rho_0)$&30MeV&$x_1$&-0.332282\\
$L$&60MeV&$x_2$&1.35830\\
$G_S$&132MeV $fm^{5}$&$x_3$&-0.0228181\\
$G_V$&$5MeV fm^{5}$&$\alpha$&0.235879\\
$G'_0(\rho_0)$&0.42 &$W_0$&133.3 MeV $fm^{5}$\\
\noalign{\smallskip}\hline
\noalign{\smallskip}\hline
\end{tabular}
\end{table*}

\begin{center}
\centering\includegraphics[width=9.0cm]{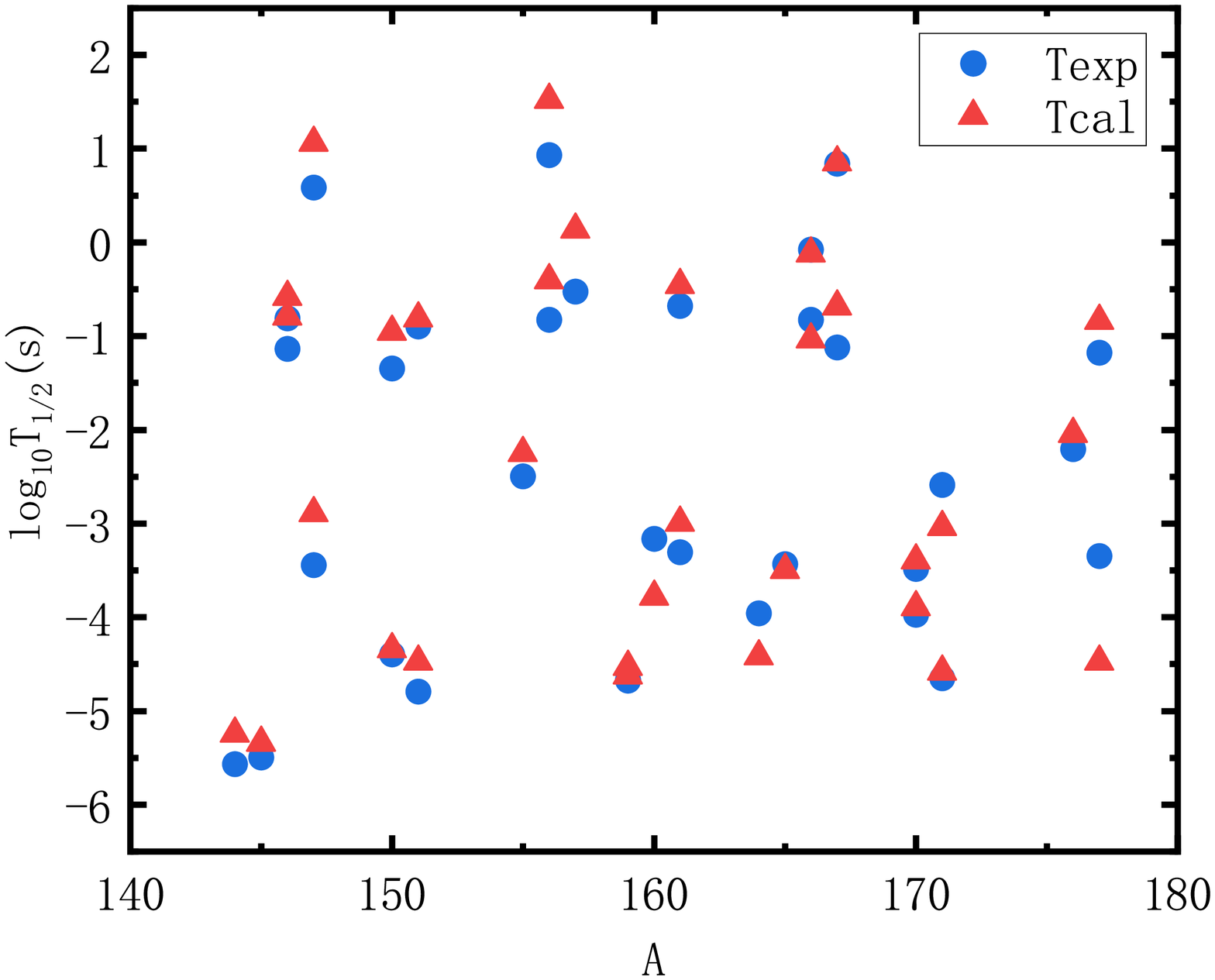}
\figcaption{(Color online) The comparison of experimental proton radioactivity half-lives and theoretical proton radioactivity half-lives calculated by TPA-SHF-MSL0 ($69\leq Z \leq81$). }
\label{fig 1}
\end{center}

Using the MSL0 parameter set, we systematically calculate the proton radioactivity half-lives of $69\leq Z \leq81$ nuclei with $S_p=1$ \cite{cheng2020systematic}. The detailed results are plotted in Fig. \ref{fig 1}. The blue circle and red triangle represent the logarithmic form of the experimental data and the logarithmic form of the calculated proton radioactivity half-life, respectively. As can be seen from Fig. \ref{fig 1}, the proton radioactivity half-lives vary in a wide range from $10^{-6}$ s to $10^2$s. Although the proton radioactivity half-life variation is up to eight orders of magnitude, the theoretical point follows the experimental point well and almost coincides with it. Furthermore, we can obtain the standard deviation $\Delta$ between the experimental data, and the calculated proton radioactivity half-life is equal to 0.405, which means the theoretical proton radioactivity half-lives calculated by TAP-SHF with MSL0 (TPA-SHF-MSL0) can reproduce the experimental data well.

To clearly reveal the dependence of the theoretical proton radioactivity half-lives on each macroscopic quantity, we varied one portion at a time while keeping all other macroscopic quantities at their default values in MSL0. In the present work, the variation of each macroscopic quantity was controlled within the experimentally obtained values to ensure that the dependencies we obtained were meaningful. To intuitively reflect the relationship between the theoretical proton radioactivity half-life change and the evolution of each macroscopic quantity, we plotted the standard deviations between experimental data and the calculated proton radioactivity half-lives corresponding to the change of each macroscopic quantity in Fig. \ref{fig 2}. From this figure, we can clearly see that the variety of different macroscopic quantities causes different effects on the standard deviations between experimental data and the calculated proton radioactivity half-lives. The standard deviation shows a robust correlation with $\rho_0$ and a relatively weak correlation with $G_S$. For the other seven macroscopic quantities, their changes have little effect on the standard deviations between experimental data and the calculated proton radioactivity half-lives. The discovery of this relationship means that we can constrain the Skyrme parameters using two macroscopic quantities that significantly affect the standard deviation, which has a higher weight in the standard deviation calculation. Thus, a new set of Skyrme parameters suitable for calculating the proton radioactivity half-life is obtained with as few adjustable parameters as possible. Similarly, such a relationship, in turn, limits the range of these two macro parameters. Its means that the actual value of $\rho_0$ is around 0.156$fm^{-3}$, and the real value of $G_S$ is about 187 MeV $fm^{5}$.

\begin{center}
\centering\includegraphics[width=6.2cm]{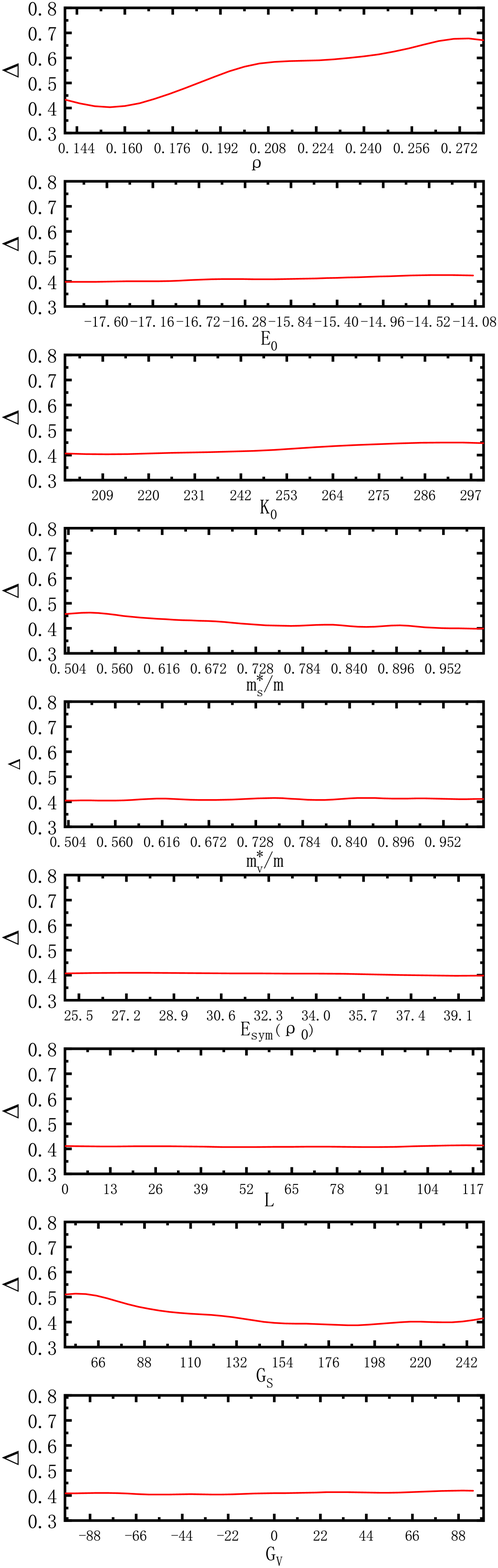}
\figcaption{(Color online) The standard deviations $\Delta$ between experimental data and theoretical proton radioactivity half-lives calculated by varying individually $\rho_0$, $E_0(\rho_0)$, $K_0$, $m^*_{s,0}$, $m^*_{v,0}$, $E_{sym}(\rho_0)$, $L$, $G_S$ and $G_V$. }
\label{fig 2}
\end{center}

In this work, the new Skyrme parameter MQSP is given by fitting the two highest weight macroscopic quantities $\rho_0$ and $G_S$, and the detailed results are listed in Tab. \ref{table 2}. The new theoretical value of the proton radioactivity half-life can be calculated by TAP-SHF with MQSP (TPA-SHF-MQSP). To verify the accuracy of our given new Skyrme parameters for calculating the proton radioactivity half-life, the universal decay law for proton radioactivity (UDLP) was chosen as a comparison. The UDLP is given by Qi \emph{et al.} \cite{PhysRevC.85.011303}, which is an extension of the universal decay law (UDL) \cite{PhysRevLett.103.072501, PhysRevC.80.044326}. In UDLP, the logarithm of proton radioactivity half-life can be expressed as 
\begin{equation}
log_{10}T_\frac{1}{2}=a\chi'+b\rho'+c+dl(l+1), 
\end{equation}
where $\chi'={Z_p}Z_d\sqrt{\frac{\mu}{Q_p}}$ and $\rho'=\sqrt{\mu{Z_p}Z_d({A_d}^\frac{1}{3}+{A_p}^\frac{1}{3})}$. Here the parameters $a=0.386, b=-0.502$, $c=-17.8$ and $d=2.386$ are determined by fitting to experimental data of proton radioactivity taken from Ref. \cite{PhysRevC.85.011303}. As a comparison, we used TPA-SHF-MSL0,TPA-SHF-MQSP and UDLP to calculate the proton radioactivity half-life, respectively, and the results are listed in Tab. \ref{table 3}. In this table, the first four columns represent the parent nuclei, the orbital angular momentum $l$ taken away by the emitted proton, the proton radioactivity energy $Q_p$, and the logarithmic form of the experimental proton radioactivity half-lives, respectively. The following three columns represent the logarithmic form of theoretical proton radioactivity half-lives calculated by TPA-SHF with the Skyrme effective interaction of MSL0 (TPA-SHF-MSL0), TPA-SHF with the Skyrme effective interaction of MQSP(TPA-SHF-MQSP), and UDLP denoted as MSL0, MQSP, and UDLP, respectively. From this table, we can clearly see that the theoretical proton radioactivity half-lives calculated by TPA-SHF-MQSP can reproduce the experimental data well.

\begin{table*}[!hbt]
\centering
\caption{Macroscopic quantities and corresponding Skyrme parameters obtained by MQSP.}
\label{table 2}
\renewcommand\arraystretch{1.5}
\setlength{\tabcolsep}{2.5pt}
\setlength\LTleft{-13in}
\setlength\LTright{-13in plus 1 fill}
\begin{tabular}{cccc}
\hline
\multicolumn{1}{c}{Quantity}&MQSP&$\rm{Quantity}$&MQSP\\
\hline
$\rho_0$&0.155 $fm^{-3}$&$t_0$&-2134.31 MeV $fm^{5}$\\
$E_0(\rho_0)$&-16MeV&$t_1$&529.92 MeV $fm^{5}$\\
$K_0$&230MeV&$t_2$&-187.13 MeV $fm^{5}$\\
$m^*_{s,0}/m$&0.8&$t_3$&13130.28 MeV $fm^{3+3\alpha}$\\
$m^*_{v,0}/m$&0.7&$x_0$&-0.0561957\\
$E_{sym}(\rho_0)$&30MeV&$x_1$&-0.3721052\\
$L$&60MeV&$x_2$& 0.159008\\
$G_S$&182MeV $fm^{5}$&$x_3$&-0.21680\\
$G_V$&$5MeV fm^{5}$&$\alpha$&0.24205\\
$G'_0(\rho_0)$&0.42 &$W_0$&133.3 MeV $fm^{5}$\\
\noalign{\smallskip}\hline
\noalign{\smallskip}\hline
\end{tabular}
\end{table*}

\begin{table*}[!hbt]
\centering
\caption{The calculation of the spherical proton radioactivity half-lives. Measured, $lgT^{\rm{MSL0}}_{1/2}$, $lgT^{\rm{MQSP}}_{1/2}$ and $lgT^{\rm{UDLP}}_{1/2}$ are the logarithmic form of the experimental proton radioactivity half-life and the calculations by STPA-SHF-MSL0, TPA-SHF-MQSP, and UDLP.}
\label{table 3}
\renewcommand\arraystretch{1.05}
\setlength{\tabcolsep}{10pt}
\setlength\LTleft{-10in}
\setlength\LTright{-10in plus 1 fill}
\begin{tabular}{cccccccc}
\hline\noalign{\smallskip}
\hline\noalign{\smallskip}
Nucleus &$l$&$Q_{p}(\rm{MeV})$ &Measured& $lgT^{\rm{MSL0}}_{1/2}$(s)& $lgT^{\rm{MQSP}}_{1/2}$(s)& $lgT^{\rm{UDLP}}_{1/2}$(s) \\
\noalign{\smallskip}\hline\noalign{\smallskip}
$^{144}$Tm&5&1.725&-5.569&-5.243&-5.42&-4.691\\
$^{145}$Tm&5&1.736&-5.499&-5.339&-5.496&-4.767\\
$^{146}$Tm$^m$&5&1.206&-1.137&-0.788&-0.958&-0.814\\
$^{146}$Tm&0&0.896&-0.81&-0.583&-0.746&-0.482\\
$^{147}$Tm&5&1.059&0.587&1.063&0.849&0.772\\
$^{147}$Tm$^m$&2&1.12&-3.444&-2.89&-3.207&-2.713\\
$^{150}$Lu$^m$&2&1.29&-4.398&-4.341&-4.52&-3.91\\
$^{150}$Lu&5&1.27&-1.347&-0.954&-1.116&-0.988\\
$^{151}$Lu$^m$&2&1.301&-4.796&-4.473&-4.647&-4.025\\
$^{151}$Lu&5&1.255&-0.896&-0.81&-0.98&-0.863\\
$^{155}$Ta&5&1.453&-2.495&-2.248&-2.39&-2.17\\
$^{156}$Ta&2&1.02&-0.826&-0.401&-0.498&-0.417\\
$^{156}$Ta$^m$&5&1.11&0.933&1.522&1.361&1.129\\
$^{157}$Ta&0&0.935&-0.527&0.137&-0.005&0.123\\
$^{159}$Re$^m$&5&1.801&-4.665&-4.531&-4.706&-4.181\\
$^{159}$Re&5&1.816&-4.678&-4.62&-4.806&-4.268\\
$^{160}$Re&0&1.267&-3.163&-3.778&-3.942&-3.408\\
$^{161}$Re$^m$&5&1.317&-0.678&-0.455&-0.633&-0.611\\
$^{161}$Re&0&1.197&-3.306&-2.992&-3.126&-2.689\\
$^{164}$Ir&5&1.844&-3.959&-4.418&-4.594&-4.114\\
$^{165}$Ir$^m$&5&1.711&-3.433&-3.496&-3.68&-3.306\\
$^{166}$Ir&2&1.152&-0.824&-1.034&-1.169&-1.014\\
$^{166}$Ir$^m$&5&1.332&-0.076&-0.118&-0.283&-0.335\\
$^{167}$Ir&0&1.07&-1.12&-0.682&-0.804&-0.645\\
$^{167}$Ir$^m$&5&1.245&0.842&0.855&0.688&0.523\\
$^{170}$Au&2&1.472&-3.487&-3.891&-4.186&-3.716\\
$^{170}$Au$^m$&5&1.752&-3.975&-3.394&-3.569&-3.234\\
$^{171}$Au$^m$&5&1.702&-2.587&-3.037&-3.205&-2.915\\
$^{171}$Au&0&1.448&-4.652&-4.578&-4.728&-4.157\\
$^{176}$Tl&0&1.265&-2.208&-2.041&-2.162&-1.919\\
$^{177}$Tl$^m$&5&1.963&-3.346&-4.473&-4.27&-4.206\\
$^{177}$Tl&0&1.172&-1.178&-0.836&-0.982&-0.863\\
											
\noalign{\smallskip}\hline
\noalign{\smallskip}\hline
\end{tabular}
\end{table*}

To intuitively compare the TPA-SHF-MSL0, TPA-SHF-MQSP, and UDLP with experimental data, the logarithmic deviations between experimental data of proton radioactivity half-lives and calculated ones are shown in Fig. \ref{fig 3}. In this figure, the X-axis represents the mass number of proton radioactivity parent nucleus, and the Y-axis represents the logarithmic deviations. The three different colors and symbols refer to the calculations obtained by three different models. From Fig. \ref{fig 3}, we can clearly see that the values of $lg(T^{cal}_{1/2}-T^{exp}_{1/2})$ obtained by TPA-SHF-MQSP are mainly near zero, indicating that the theoretical half-life of proton radioactivity calculated using our model is in good agreement with the experimental data. Furthermore, this figure shows that the calculations using TPA-SHF-MQSP can better reproduce the experimental data than TPA-SHF-MSL0 and UDLP for most nuclei. In order to intuitively compare the deviations between the half-lives of proton radioactivity obtained by different models and experimental data, we calculated the standard deviation and obtained $\Delta_{MSL0}=0.405$, $\Delta_{MQSP}=0.362$, and $\Delta_{UDLP}=0.459$ represent standard deviations between $lgT^{MSL0}_{1/2}$, $lgT^{MQSP}_{1/2}$, $lgT^{UDLP}_{1/2}$, and$lgT^{exp}_{1/2}$, respectively. These results show that the TPA-SHF-MQSP method is better than the other models in calculating the spherical proton radioactivity half-lives. Therefore, it is credible to use TPA-SHF-MQSP to study the proton radioactivity half-lives.

\begin{center}
\centering\includegraphics[width=9.0cm]{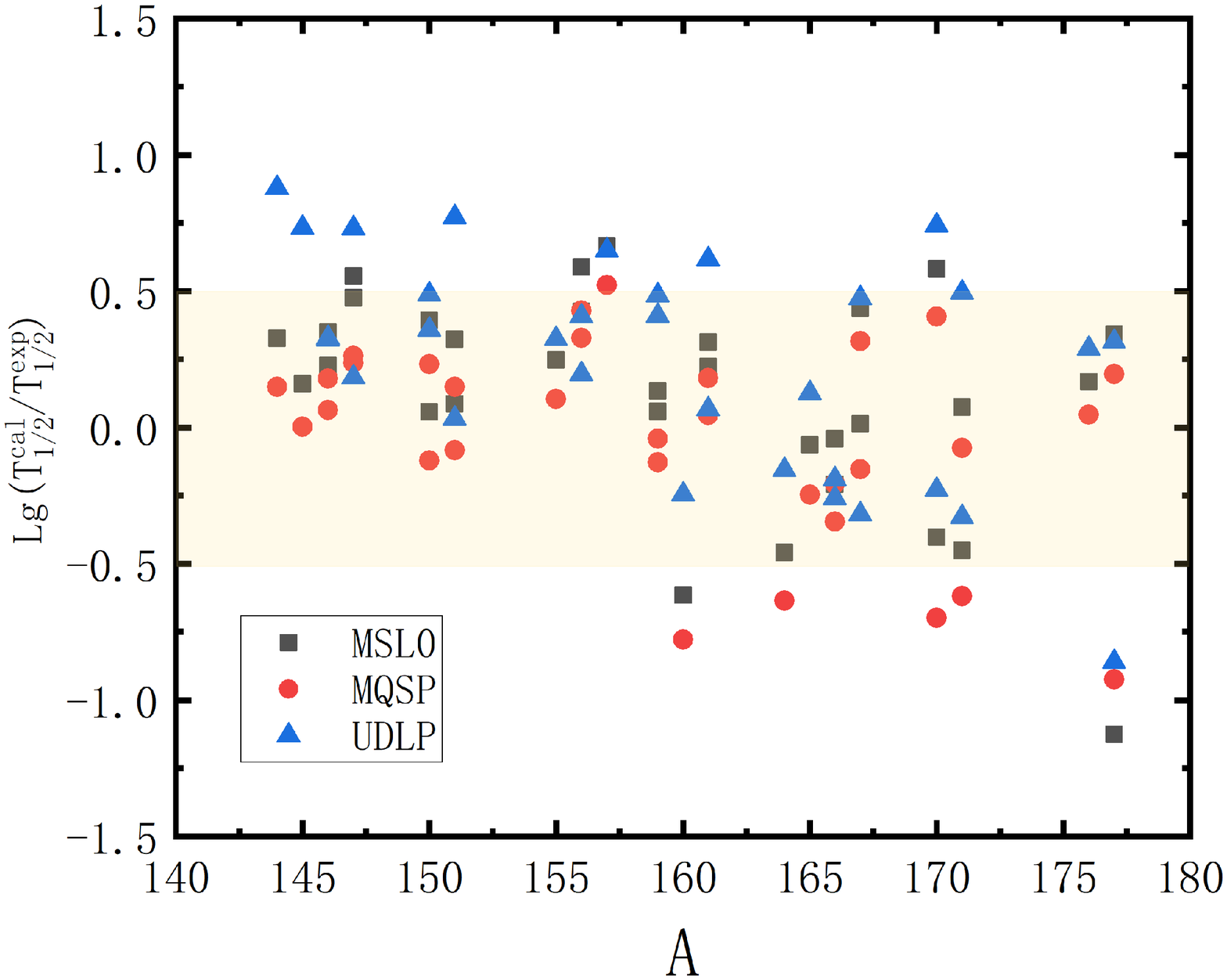}
\figcaption{(Color online) The deviations between $lgT^{cal}_{1/2}$ and $lgT^{exp}_{1/2}$ . The black square, the red circle, and the blue triangle represent the calculation results obtained using TPA-SHF-MSL0, TPA-SHF-MQSP, and UDLP, respectively. }
\label{fig 3}
\end{center}

\section{Summary}
\label{section 4}
In summary, we have systematically studied the proton radioactivity half-life of 33 spherical nuclei by investigating the relationship between the macroscopic quantities of nuclear matter and Skyrme parameters. The calculated results indicate that the proton radioactivity half-lives and $\rho_0$ showed a robust correlation, while the correlation with $G_S$ is relatively weak. For the other seven macroscopic quantities, their variations have little effect on the evolution of the standard deviations between experimental data and the calculated proton radioactivity half-lives. This effectively constrains the range of $\rho_0$ and $G_S$. Moreover, we obtain new Skyrme parameters MQSP by fitting $\rho_0$ and $G_S$, the two most weighted macroscopic quantities. Compared with Skyrme parameters MSL0 and UDLP, the theoretical proton radioactivity half-life calculated by Skyrme parameters MQSP can better reproduce the experimental data. This work can be used as a reference for future research of proton radioactivity.
\end{multicols}
\vspace{-1mm}
\centerline{\rule{80mm}{0.1pt}}
\vspace{2mm}

\begin{multicols}{2}

\end{multicols}

\clearpage
\end{CJK*}

\begin{thebibliography}{90}
\vspace{3mm}
\bibitem{SONZOGNI20021} A. A. Sonzogni, Nucl. Data Sheets \textbf{95}: 1 (2002). 
\bibitem{JACKSON1970281} K. P. Jackson, C. U. Cardinal, H. C. Evans $et \ al. $, Phys. Lett. B \textbf{33}: 281 (1970). 
\bibitem{KARNY200852} M. Karny, K. P. Rykaczewski, R. K. Grzywacz $et \ al. $, Phys. Lett. B \textbf{664}: 52 (2008). 
\bibitem{SONZOGNI2002837} A. A. Sonzogni, Nucl. Data Sheets \textbf{95}: 837 (2002). 
\bibitem{PhysRevC.69.054311} A. T. Kruppa and W. Nazarewicz, Phys. Rev. C \textbf{69}: 054311 (2004). 
\bibitem{FERREIRA2007418} L. S. Ferreira, M. C. Lopes and E. Maglione, Prog. Part. Nucl. Phys. \textbf{59}: 418 (2007). 
\bibitem{PhysRevC.79.054330} J. M. Dong, H. F. Zhang and G. Royer, Phys. Rev. C \textbf{79}: 054330 (2009). 
\bibitem{2009-0067} J. M. Dong, H. F. Zhang, W. Zuo $et \ al. $, Chin. Phys. C \textbf{34}: 2009 (2010). 
\bibitem{Hong_Fei_2009} H. F. Zhang, J. M. Dong, Y. Z. Wang $et \ al. $, Chin. Phys. Lett. \textbf{26}: 072301 (2009). 
\bibitem{Routray2012} T. R. Routray, A. Mishra, S. K. Tripathy $et \ al. $, Eur. Phys. J. A \textbf{48}: 77 (2012). 
\bibitem{DELION2006113} D. S. Delion, R. J. Liotta and R. Wyss, Phys. Reports \textbf{424}: 113 (2006). 
\bibitem{PhysRevC.58.3011} S. \AA{}berg, P. B. Semmes and W Nazarewicz, Phys. Rev. C \textbf{58}: 3011 (1998). 
\bibitem{FERREIRA2011508} L. S. Ferreira, E. Maglione and P. Ring, Phys. Lett. B \textbf{701}: 508 (2011). 
\bibitem{doi:10.1063/1.2827286} J. S. Al-Khalili, A. J. Cannon and P. D. Stevenson, AIP Conf. Proc. \textbf{961}: 66 (2007). 
\bibitem{PhysRevC.5.626} D. Vautherin and D. M. Brink, Phys. Rev. C \textbf{5}: 626 (1972). 
\bibitem{Zdeb2016} A. Zdeb, M. Warda, C. M. Petrache $et \ al. $, Eur. Phys. J. A \textbf{52}: 323 (2016). 
\bibitem{Chen_2019} J. L. Chen, X. H. Li, J. H. Cheng, $et \ al. $, J. Phys. G Nucl. Part. Phys. \textbf{46}: 065107 (2019). 
\bibitem{PhysRevLett.59.262} S. A. Gurvitz and G. Kalbermann, Phys. Rev. Lett. \textbf{59}: 262 (1987). 
\bibitem{Qian_2010} Q. Tang and X. Y. Wang, Chin. Phys. Lett. \textbf{27}: 030508 (2010). 
\bibitem{PhysRevC.72.051601} D. N. Basu, P. Chowdhury and C. Samanta, Phys. Rev. C \textbf{72}: 051601 (2005). 
\bibitem{Deng2019} J. G. Deng, X. H. Li, J. L. Chen $et \ al. $, Eur. Phys. J. A \textbf{55}: 58 (2019). 
\bibitem{PhysRevC.96.034619} K. P. Santhosh and I. Sukumaran, Phys. Rev. C \textbf{96}: 034619 (2017). 
\bibitem{CHENG2020121717} J. H. Cheng, J. L. Chen, J. G. Deng $et \ al. $, Nucl. Phys. A \textbf{997}: 121717 (2020). 
\bibitem{cheng2020systematic} J. H. Cheng, X. Pan, Y. T. Zou $et \ al. $, Eur. Phys. J. A \textbf{56}: 273 (2020). 
\bibitem{PhysRevC.82.024321} L. W. Chen, C. M. Ko, B. A. Li $et \ al. $, Phys. Rev. C \textbf{82}: 024321 (2010). 
\bibitem{doi:10.1063/1.531270} Morehead and J. James, J. Math. Phys. \textbf{36}: 5431 (1995). 
\bibitem{PhysRevC.82.059902} V. Y. Denisov and A. A. Khudenko, Phys. Rev. C \textbf{82}: 059902 (2010). 
\bibitem{Royer_2000} G. Royer, J. Phys. G Nucl. Part. Phys. \textbf{26}: 1149 (2000). 
\bibitem{PhysRevC.98.054614}Z. Zhang and C. M. Ko, Phys. Rev. C \textbf{98}: 054614 (2017). 
\bibitem{PhysRevC.80.014322} L. W. Chen, B. J. Cai, C. M. Ko $et \ al. $, Phys. Rev. C \textbf{80}: 014322 (2009). 
\bibitem{BORZOV2006645}I. N. Borzov, Nucl. Phys. A \textbf{777}: 645 (2006). 
\bibitem{PhysRevC.72.014310} B. K. Agrawal, S. Shlomo and V. K. Au, Phys. Rev. C \textbf{72}: 014310 (2005). 
\bibitem{PhysRevC.72.067303} T. Wakasa, M. Ichimura and H. Sakai, Phys. Rev. C 72: 067303 (2005). 
\bibitem{NUBASE2020} F. G. Kondev, M. Wang, W. J. Huang $et \ al. $, Chin. Phys. C \textbf{45}: 030001 (2021). 
\bibitem{CPC-2021-0034}W. J. Huang, M. Wang, F. G. Kondev $et \ al. $, Chin. Phys. C \textbf{45}: 030002 (2021). 
\bibitem{CPC-2020-0033}M. Wang, W. J. Huang, F. G. Kondev $et \ al. $, Chin. Phys. C \textbf{45}: 030003 (2021). 
\bibitem{BLANK2008403} B. Blank and M. Borge, Prog. Part. Nucl. Phys. \textbf{60}: 403 (2008). 
\bibitem{PhysRevC.85.011303} C. Qi, D. S. Delion, R. J. Liotta $et \ al. $, Phys. Rev. C \textbf{95}: 011303 (2012). 
\bibitem{PhysRevLett.103.072501} C. Qi, F. R. Xu, R. J. Liotta $et \ al. $, Phys. Rev. Lett. \textbf{103}: 072501 (2009). 
\bibitem{PhysRevC.80.044326} C. Qi, F. R. Xu, R. J. Liotta, Phys. Rev. C \textbf{95}: 044326 (2009). 
\end{thebibliography}
\end{document}